%% file: main.tex
\documentclass{article}

\usepackage{arxiv}

\usepackage[utf8]{inputenc} 

\usepackage{amsfonts}       
\usepackage{nicefrac}       
\usepackage{microtype}      
\usepackage{lipsum}

\usepackage{color}
\usepackage{listings}
\usepackage{booktabs}
\usepackage{array}
\usepackage{graphicx}
\usepackage{longtable}
\usepackage{subfigure}

\usepackage{cite}
\usepackage{url}
\usepackage{fancyhdr}

\usepackage{soul}

\usepackage{float}
\usepackage{listings}
\usepackage{mdwmath}
\usepackage{mdwtab}
\usepackage{multirow}
\usepackage{multicol}
\usepackage{rotating}
\usepackage{setspace}
\usepackage[utf8]{inputenc}
\usepackage{lineno}
\usepackage{listings}

\usepackage{mdwmath}
\usepackage{mdwtab}
\usepackage{multirow}
\usepackage{multicol}
\usepackage{array}
\usepackage{booktabs}

\usepackage{enumitem}
\usepackage{xspace}
\usepackage[export]{adjustbox}
\usepackage{graphicx}
\usepackage{color,soul}
\usepackage{rotating}
\usepackage{setspace}
\usepackage{amsmath} 
\usepackage{amssymb}
\usepackage{float}
\usepackage{xcolor}

\usepackage{hyperref}
\usepackage[numbers]{natbib}
\usepackage{url}

\title{Towards a Roadmap on Software Engineering for Responsible AI}

\author{Qinghua Lu, Liming Zhu, Xiwei Xu, Jon Whittle, Zhenchang Xing\\
Data61, CSIRO, Australia\\
firstname.lastname@data61.csiro.au}

\begin{document}

\maketitle

\begin{abstract}

Although AI is transforming the world, there are serious concerns about its ability to behave and make decisions responsibly. Many ethical regulations, principles, and frameworks for responsible AI have been issued recently. However, they are high level and difficult to put into practice. On the other hand, most AI researchers focus on algorithmic solutions, while the responsible AI challenges actually crosscut the entire engineering lifecycle and components of AI systems. To close the gap in operationalizing responsible AI, this paper aims to develop a roadmap on software engineering for responsible AI. The roadmap focuses on 
(i) establishing multi-level governance for responsible AI systems, (ii) setting up the development processes incorporating process-oriented practices for responsible AI systems, and (iii) building responsible-AI-by-design into AI systems through system-level architectural style, patterns and techniques.

\end{abstract}


\section{Introduction}
\label{intro}

Artificial intelligence (AI) is considered one of the major driving forces to transform society and industry and has been successfully adopted in 
data-rich domains. 
Although AI is solving real-world challenges and improving our quality of life, there are serious concerns about its ability to behave and make decisions in a responsible manner.
Responsible AI has become one of the greatest scientific challenges of our time. Both legal and ethical aspects need to be considered to achieve responsible AI. Since the law establishes the minimum standards of behaviour while ethics sets the maximum standards, in this paper, we use the terms \textit{responsible AI}, \textit{ethical AI} and \textit{ethics} to cover the broader set of requirements. 

A large number of ethical principle frameworks have been recently issued by governments, research institutions, and enterprises ~\cite{jobin2019global}, which responsible AI technologies and systems are supposed to adhere to. 
A degree of consensus around the principles has been achieved~\cite{fjeld2020principled}. 
However, these principles are 
high-level and do not provide operationalized guidance and software engineering methods on how to develop responsible AI systems. This leaves unanswered questions such as how can these principles be designed for, implemented and tracked in developing and operating an AI system. 
On the other hand, significant research has gone into ethical algorithms where the formulation of some ethical principles is amenable to mathematical definitions, analysis and theoretical guarantees. These algorithm-level mechanisms mainly focus on a small subset of ethical principles (such as privacy~\cite{ji2014differential} and fairness~\cite{mehrabi2019survey}) relying on theoretical heuristics. There is a lack of linkage to the software development processes, especially requirements engineering, system design methods, or operations.

\begin{figure}
\centering
\includegraphics[width=0.5\columnwidth]{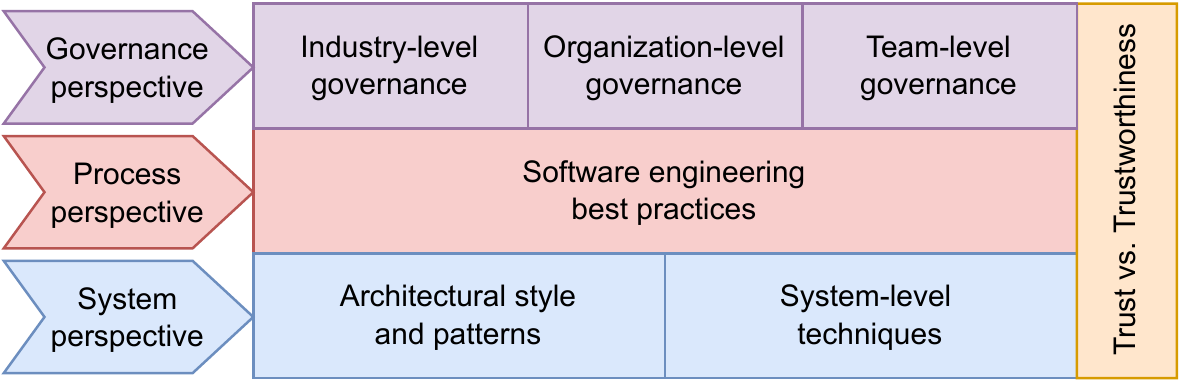}
\caption{Overview of the roadmap.} \label{fig:overview}
\vspace{-2ex}
\end{figure}

Therefore, this paper presents a research roadmap on software engineering for operationalizing responsible AI. Rather than staying at the ethical principle level or going straight down to the AI algorithm level, this paper focuses on the software engineering approach to operationalizing responsible AI. We perform a systematic literature review (SLR) on software engineering for responsible AI to summarize the current state and identify the critical research challenges. As shown in Fig.~\ref{fig:overview}, the proposed roadmap focuses on (i) establishing multi-level governance for responsible AI systems, (ii) setting up the development processes incorporating process-oriented best practices for responsible AI systems, and (iii) building responsible-AI-by-design into the AI systems through system-level architectural style, patterns and techniques. 

The remainder of the paper is organized as follows. Sec. 2 discusses the methodology. The rest of the paper is divided into three parts: governance perspective (Sec. 3), process perspective (Sec. 4), and system perspective (Sec. 5). For each perspective, we present the current state and the challenges being faced by the community.


\section{Methodology}
\begin{figure*}[b!]
\centering
\includegraphics[width=0.7\textwidth]{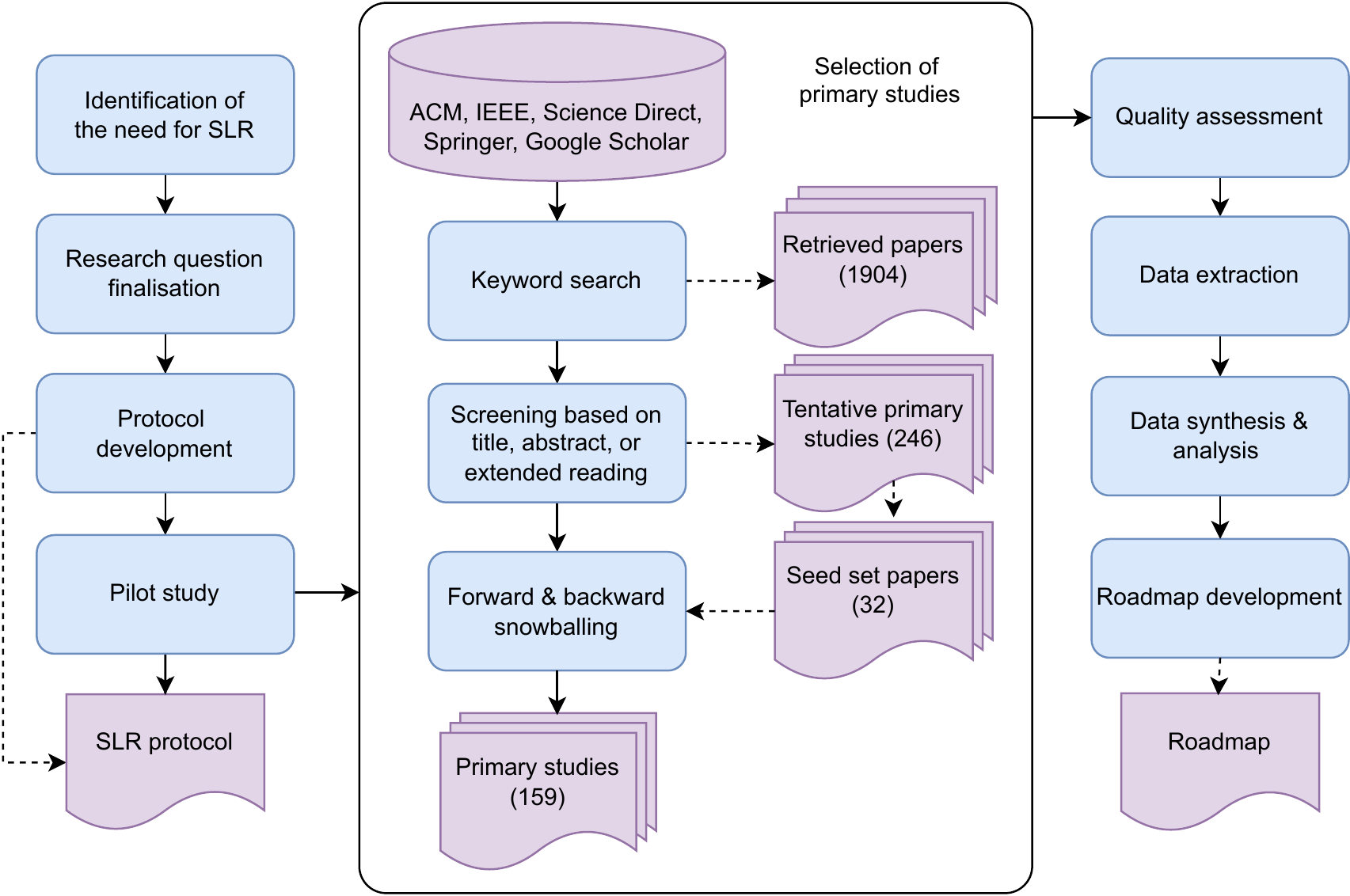}
\caption{Methodology.} \label{fig:methodology}
\vspace{-2ex}
\end{figure*}

To develop a roadmap, we performed an SLR following the guideline in ~\cite{keele2007guidelines}. Fig.~\ref{fig:methodology} illustrates the methodology.
The two research questions are defined for the SLR: 1) What responsible AI principles are addressed by the study;
2) What solutions for responsible AI can be identified.
The data sources include ACM Digital Library, IEEE Xplore, Science Direct, Springer Link, and Google Scholar. 
The study only includes papers presenting concrete solutions for responsible AI and exclude papers discussing high-level frameworks.
A set of 159 primary studies was identified. The complete SLR protocol is available as online material \footnote{\url{https://drive.google.com/file/d/16fawGwzuuMwFpCAl4mH-MtNcWQRpFNtv/view?usp=sharing}}. We use the ethical principles listed in Harvard University's mapping study ~\cite{fjeld2020principled}. Fig.~\ref{fig:principles_summary} lists an adapted summary of the principles (responsibility is merged into accountability given the overlapping definitions).

\begin{figure}[b!]
\vspace{-1ex}
\centering
\fbox{%
\centering
\begin{minipage}{0.96\columnwidth}
\small
\begin{enumerate}[leftmargin=*]
\item
{\bf Privacy.}
AI systems should preserve the data privacy.

\item
{\bf Accountability.}
Those responsible for the various phases of the AI system lifecycle should be identifiable and accountable
for the outcomes of the system.

\item
{\bf Safety \& Security.}
AI systems should safely and securely operate in accordance with their intended purpose throughout their lifecycle.

\item
{\bf Transparency \& Explainability.}
There should be responsible disclosure to ensure people know when they are being engaged/impacted by an AI system. The behaviors and decisions should be explainable.

\item
{\bf Fairness and Non-discrimination.}
AI systems should be inclusive and should not involve or result in unfair discrimination
against individuals, communities or groups.

\item
{\bf Human Control of Technology.}
When an AI system impacts a person, community, or environment, there should be a timely process to allow people to challenge the use or output of the system.

\item
{\bf Promotion of Human Values.}
AI systems should respect human rights, diversity, and the autonomy of individuals, and benefit individuals, society and the environment.
\end{enumerate}
\end{minipage}
} 
\vspace{-1ex}
\caption
  {
  An adapted summary of AI Ethics Principles~\cite{fjeld2020principled}.
  }
\label{fig:principles_summary}
\end{figure}

\section{Governance Perspective}

\begin{figure*}
\centering
\includegraphics[width=0.85\textwidth]{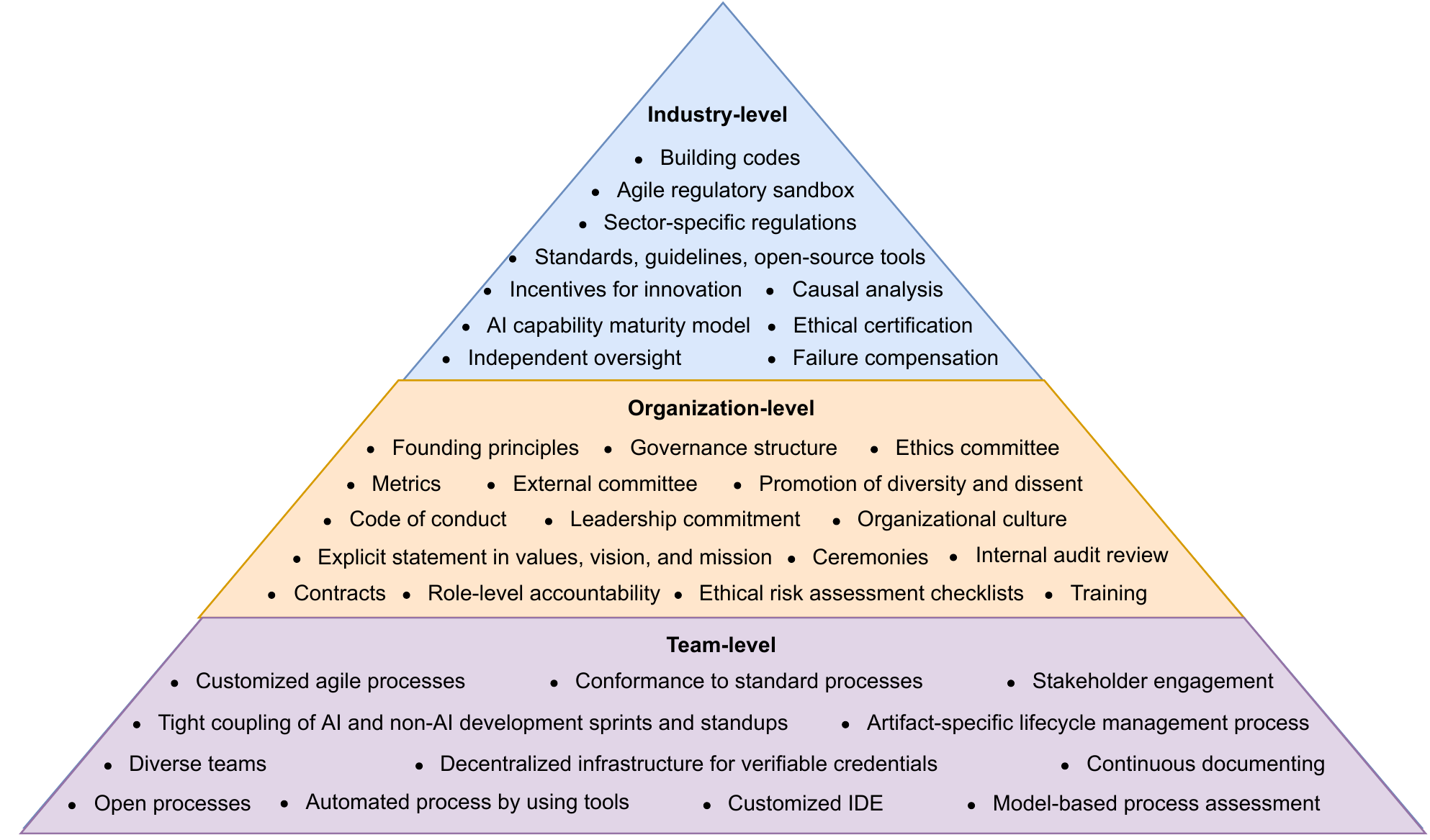}
\caption{Multi-level governance for responsible AI.} \label{fig:governance}
\vspace{-2ex}
\end{figure*}

\subsection{Current State}
The governance for responsible AI systems can be defined as the structures and processes that are designed to ensure the development and use of AI systems are compliant to ethical regulations and responsibilities. As shown in Fig.~\ref{fig:governance}, the governance can be built into three levels based on Shneiderman's governance structure~\cite{Shneiderman20}: industry-level, organization-level, and team-level.

\subsubsection{Industry-level governance} \noindent \par

The industry-level governance requires the governments and AI industry to act collectively via regulation, policy, standards to make the AI systems acceptable by society. One form of regulation is analogous to \textbf{building codes} to which developers can adhere when developing AI systems~\cite{Landwehr15}. \textbf{Incentives/penalties} may be applied for ethical/unethical software development~\cite{Shneiderman20}. Given the laws and public policies usually take a long time to enact, governments can consider adopting \textbf{agile regulatory sandbox} on a time-limited basis for the emerging AI technology. For example, the Singapore government and EU have applied a regulatory sandbox to allow autonomous vehicles to be legally on the roads without changing the national laws~\cite{Borg21}. As there may be various sector/domain-specific risk concerns (e.g. military or health), there is a need to extend and adapt generic regulations to \textbf{sector/domain-specific regulations}~\cite{Ibanez21}. Professional and non-governmental organizations and research institutions have been making significant efforts on developing \textbf{standards, guidelines, and open-source tools and platforms}\cite{Shneiderman20}. Regulators could \textbf{incentivize organizations for responsible AI innovation}. AI project funding bodies could require applicants to include \textbf{responsible AI statements} in their funding applications and implement \textbf{ethical checklist driven monitoring and management of grant funding}.

\textbf{Independent oversight} is essential to the accountability of AI systems. 
\textbf{External audits} could be conducted by independent third parties (such as AI Safety Commission) during the development or post-hoc.
The inspection of the AI systems' behaviours and decision-making is required either by reviewing interpretable AI models or having access to the artifacts of AI systems, such as datasets, source code, and documents.
For example, Z-inspection process~\cite{Zicari21} is a generic \textbf{inspection processes} to assess the trustworthiness of AI systems.
When an accident happens, \textbf{causal analysis} can be performed using the \textbf{why-because} method for accident investigation. Insurance companies can play a role of a guarantor for responsible AI and \textbf{compensate for the failures} of AI systems
~\cite{Shneiderman20}.

\textbf{AI capability maturity model} is being introduced to examine organizations' AI capability ~\cite{Alsheibani2019,Fukas2021}. Like the conventional software engineering capability maturity model, the AI capability maturity model has different levels of maturity based on development processes and desired responsible AI metrics. The results assessed by the AI capability maturity model could be used for \textbf{ethical certification}~\cite{Cihon21} to certify an organization's ability to achieve ethical principles. In addition to organization-level assessment and certification, both the AI capability maturity model and ethical certification could be extended to cover a broader view to provide verifiable evidence for improving human trust in AI systems, e.g., assessing and certifying for organizations, development processes, developers, operators, AI systems, components, models.

\subsubsection{Organization-level governance} \noindent \par

Achieving responsible AI in an organization requires the \textbf{establishment of AI governance} that includes setting up \textbf{founding principles}, \textbf{an ethics committee}, \textbf{a governance structure}, \textbf{governance metrics}, \textbf{external committees}, \textbf{organization-wide training}, and \textbf{promotion of diversity and dissent}~\cite{Eitel-Porter21}. The ethics committee should oversee the overall AI-driven decision-making processes (not only the AI systems). \textbf{Code of conduct} (such as specific code of ethics) should be developed 
for the employees (e.g. developers or operators) to follow~\cite{Shneiderman20}.

\textbf{Leadership commitment} is essential to organization-level governance. \textbf{Responsible AI statements} should be described explicitly in an organization's \textbf{values, vision, and mission}~\cite{Shneiderman20}. The establishment of responsible AI governance could be incorporated into \textbf{CEO's contracts and performance reviews}~\cite{Borg21}. The management can enforce the \textbf{organizational culture} on responsible AI and use \textbf{ceremonies} to celebrate responsible AI successes (e.g. certificate granted). AI transformation workshops can be organized 
to assess the impact, e.g. using \textbf{human-centered AI canvas}\footnote{\url{https://medium.com/@albmllt/introducing-the-human-centered-ai-canvas-a4c9d2fc127e}}.

\textbf{Internal audit review} is required to cover the complete lifecycle of AI systems and include continuous monitoring instruments (e.g. checklist for retrospective meetings and data/code reviews). \textbf{Extensive reporting of failures and near misses} should be produced with why-because analysis. \textbf{Ethical risk assessment checklists}~\cite{Jacovi21} need to be \textbf{co-designed with practitioners}~\cite{Madaio20} (e.g. for ethics committees, team, or prospective purchasers) for each of the ethical principles (e.g. fairness risk identification questionnaire~\cite{Lee21}, reproducibility checklists, fairness checklists~\cite{Jacovi21}), \textbf{taking into account the application category and automation level} for risk and timeliness.

\textbf{Role-level accountability} ~\cite{zhu2021ai} is established in the organization through \textbf{formal contractual mechanisms} (e.g., legal agreements between system users, data contributors, and the project team) to hold each other accountable~\cite{Jackson21}. Diverse types of ethical quality constraints are enforced as a contract, such as service contract, model contract, and data contract. The provenance of data, model, and code allows examining role-level accountability.

\textbf{AI ethics training} programs should be introduced within the organizations, including \textbf{technical and non-technical ethics and human rights training for different roles (such as the management, developers, data scientists, operators) and organizational awareness}. The content of an AI ethics training course could include \textbf{governance for responsible AI}, \textbf{ethical operations of AI systems}, \textbf{trustworthy development processes}, and \textbf{responsible-AI-by-design with case studies} (e.g. using the design of an ethical/unethical agent~\cite{Weiss21}, autonomous vehicles~\cite{Heidi19}).

\subsubsection{Team-level governance} \noindent \par

Team-level governance is mainly for managing the AI projects and overall development processes. The development processes should be \textbf{conformance to standard processes}. \textbf{Agile development processes can be adapted and customised} by incorporating responsible AI principles ~\cite{hussain2021}. \textbf{The AI and non-AI development sprints and standups need to be closely coupled}~\cite{Amershi19}. One effective way to ensure fairness, human-centred values, and HSE wellbeing in AI systems is to make the development \textbf{teams diverse} (e.g., gender, age, race, culture) and \textbf{engage stakeholders} throughout the lifecycle of AI systems. 
Culture needs to be explicitly considered in the design when there is culture-sensitive data or context involved. For example, culture could determine whether it is ethical or unethical for a conversational AI system to tell a lie to humans (e.g. when negotiating a price
)~\cite{kim21}. For indigenous projects, indigenous people need to be involved in the development process to help \textbf{incorporate culture concerns into development and decision-making} (e.g., following CARE\footnote{\url{https://www.gida-global.org/care}}). 
\textbf{Ethicists can be included into the development team} to promote a more ethical development of AI systems~\cite{McLennan2020}. Conflict/trade-off resolution process is needed to achieve consensus during the development.

The day-to-day workflow of software engineers (e.g. training and deployment pipeline~\cite{Amershi19}) is expected to be \textbf{unified and automated by using the standardized tools} (e.g., Jupiter notebooks, python, Github)~\cite{Papagiannidis21} to improve productivity and integrate AI ethics tools. Model-based process assessment checks the process compliance by \textbf{linking workflow models with assessment criteria}~\cite{Woitsch21}. \textbf{Customized visual IDE tools} may help developers with varying levels of experience~\cite{Amershi19}. The project team can consider \textbf{publishing the findings through scientific publication} following the scientific norms of research integrity and knowledge sharing to provide public transparency~\cite{Jackson21}.
\textbf{Open process across the development lifecycle} (e.g. full access to artifacts from stakeholders or authorized third-parties) and \textbf{Artifact specific lifecycle management process} are helpful to improve transparency and accountability~\cite{Hutchinson21}. 
\textbf{Decentralized computational infrastructure} is built for \textbf{verifiable ethical credentials} which can be used as proofs of responsible AI compliance~\cite{chu2021}. 
Developers are suggested to prepare standardized documentation (e.g. ISO AI standards\footnote{\url{https://www.iso.org/committee/6794475/x/catalogue/}}). There are various types of \textbf{documentation} for managing the development of responsible AI systems, such as \textbf{model cards}, \textbf{data statements}, \textbf{datasheets for datasets}, \textbf{AI service factsheets}~\cite{Jacovi21}, \textbf{requirements specification},	\textbf{design document}, \textbf{implementation diary}, \textbf{testing report}, \textbf{maintenance plan}~\cite{Hutchinson21}.

\subsection{Research Challenges}
\textbf{Establishing multi-level governance for responsible AI.}
Despite nearly one hundred frameworks for ethical principles being issued ~\cite{jobin2019global}, these principles are too high-level and hard to operationalize. 
A systematic and holistic governance framework connected with software development lifecycles is highly desirable to translate ethical principles into practices throughout the entire lifecycle of AI systems. The governance structure and processes need to be designed and organized into multiple levels, including industry-level (which can be further divided into international-level and country-level), organization-level, and team-level. So the governments, organizations and development teams can follow the framework to governance AI systems.

\textbf{Governance mechanisms linking with stakeholders and lifecycle processes.} The governance mechanisms need to be designed in a way that links with stakeholders and process stages. This makes the practitioners easy to adopt the governance mechanisms in practice. Also, to identify and track accountability, we need to understand the stakeholders of AI systems and their roles in the governance throughout the entire lifecycle of AI systems. 

\textbf{The communication between the SE people and the rest (i.e. people who are not SE-people).}
Responsible AI challenges are broader than software engineering and need to be addressed by a multidisciplinary team with expertise in software engineering, machine learning, social science, human-machine interaction, and user experience. However, as the final deliverable to the society is a responsible AI system, we believe the software engineering people are the key force to drive the research and collaboration.

\textbf{Education for students.}
As there is a serious concern about the social impact of AI systems, education on responsible AI is urgently required for students across all across all levels (such as K-12 and university education).



\begin{figure*}
\centering
\includegraphics[width=0.85\textwidth]{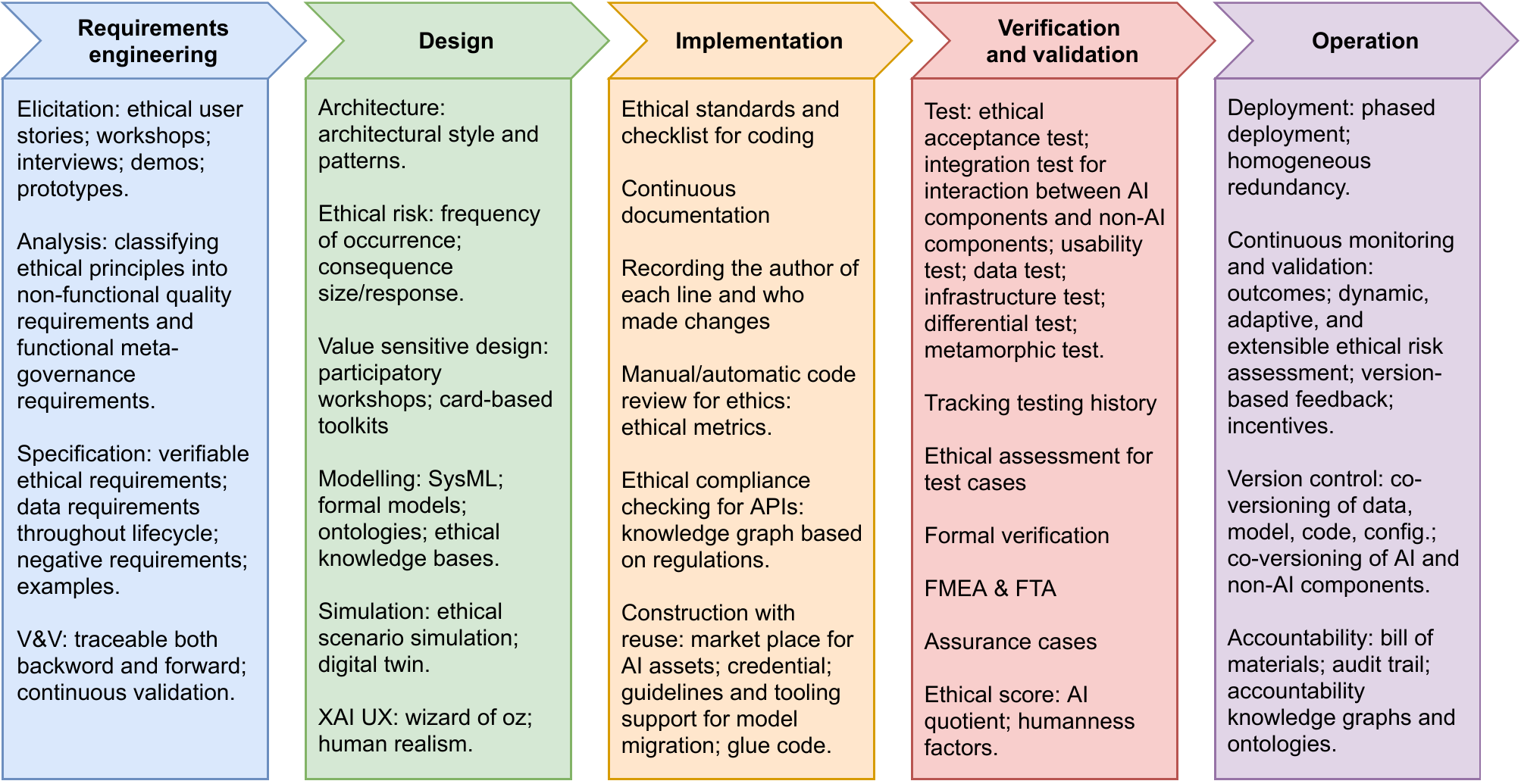}
\caption{Development process practices.} \label{fig:practices}
\vspace{-2ex}
\end{figure*}

\section{Process Perspective}

\subsection{Current State}
Fig.~\ref{fig:practices} summarises the methods and best practices that can be incorporated into development processes, so the developers could consider to apply them during the development. 

\subsubsection{Requirement engineering} \noindent \par
Responsible AI requirements are either omitted or mostly stated as high-level project objectives in practice. The existing methodologies
should be extended and adopted for AI systems to ensure that the requirements captured are as accurate and complete as possible while recognizing the special characteristics of AI systems such as autonomy, continuous learning and the value-alignment problem.

\textbf{Requirements Elicitation: }
Elicitation is the process of \textbf{collecting requirements from stakeholders} and other sources, including goals, domain knowledge, business rules, operational environment, organizational environment~\cite{SWEBOK}. Ethical principles are an essential source for identifying ethical requirements types and relevant and inclusive stakeholders. Some of the important stakeholders for collecting ethical requirements include not only business owners, system users, ethical/legal experts, regulators but also data providers,impacted data subjects, operators, advocacy groups and even concerned public. \textbf{Culture safety} is a critical ethical requirement for AI systems that involve culture sensitive data. There are a variety of requirements elicitation techniques that can be adapted to gather ethical requirements (e.g. training/validating data fairness requirements and secondary data usage requirements), such as \textbf{interviews}~\cite{JingWen18}, \textbf{scenarios}, \textbf{requirement workshops}, \textbf{interactive demos/prototypes}~\cite{JingWen18}, and \textbf{user stories}. \textbf{Ethical user stories} (e.g. utilizing ECCOLA cards~\cite{Halme21}) is an effective way to transform the abstract ethical requirements into tangible outcomes.

\textbf{Requirements Analysis: }
One of the most important activities in requirement analysis is requirements classification~\cite{SWEBOK} and resolution of requirements conflicts. We group the ethical principles (in Fig.~\ref{fig:principles_summary}) into \textbf{two requirements categories based on their nature and characteristics}.
The first group includes privacy, safety \& security, fairness, and human values, which are similar to \textbf{software qualities}~\cite{iso2011iec25010} and can be considered as \textbf{non-functional requirements}. Some principles, such as safety \& security, are the quality attributes well studied in the dependability community~\cite{dependability} and can be identified and considered early in the development lifecycle. Recurring requirement problems and associated reusable design fragments, like patterns/tactics, could be applied to meet those quality attributes~\cite{bass2003software}. Privacy is not a standard software quality attribute~\cite{iso2011iec25010}, but has been treated as an increasingly important non-functional requirement to realize regulatory requirements, like GDPR\footnote{General Data Protection Regulation (GDPR), https://gdpr-info.eu/.}. 
Recurring requirements and reusable patterns/practices have been summarized in for privacy \cite{privacyDesign}. Fairness is another requirement that the developers should collect and design for from early development life cycles. Technical mechanisms/practices to remove bias at different stages of the pipeline have been designed ~\cite{mehrabi2019survey}. There has been recently emerging research on casting human values into requirements in software engineering and their operationalization ~\cite{hussain2020casestudy,hutchinson2020accountability}, including the extension of value-based design methods 
\cite{VSD2015},  extension of human factor research on productivity and usability into human values consideration. But these efforts are still limited to a small subset of human values \cite{Whittle20-ICSE}. 
The reason to group these principles is that they can be handled and validated using how non-functional quality attributes are handled in software system design. Some principles can be quantitatively validated, whereas others can be qualitatively handled by the widely used design patterns and process-oriented practices. 

The second group includes transparency \& explainability, human control of technology, and accountability, which are \textbf{meta-level governance issues} and can be classified as \textbf{functional requirements for improving users' confidence in the AI system}. Transparency \& explainability can be fulfilled by designing functions for receiving an explanation of a prediction or decision and having access to the artifacts of AI systems. 
Human control of technology requires a function that allows the users to challenge the output or use of the AI system. To meet accountability, the AI system should include a function for tracing and identifying those who are responsible for the various phases of the AI system lifecycle and the outcomes of the system.

\textbf{Requirements Specification: }
AI systems are complex software systems that involve hardware components. Thus, both software requirements and system requirements (e.g., requiring certified hardware components) are needed. It's worth noting that AI systems try to solve the problems autonomously with a level of independence and agency and cannot be fully specified. It is important to judiciously make \textbf{ethical requirements quantifiable or measurable}, and avoid vague and unverifiable requirements~\cite{SWEBOK}. \textbf{Scope of responsibility} need to be clearly defined in the requirements~\cite{responsibility_meaning_chi21}. Both \textbf{team diversity} and the choice of \textbf{ethical requirements verification/validation techniques} can be specified as process requirements. 
\textbf{Data requirements}~\cite{Req_REW} need to be listed explicitly and specified throughout the data lifecycle (i.e., collection, management, survey, consumption, termination) taking into account all the involved roles (i.e., data provider, manager, analyser, consumer) and ethical concerns, e.g. training/validation data fairness requirements and secondary data usage requirements. \textbf{Functional requirements should be stated with performance measures}~\cite{Req_REW} and their explanations in the context of the domain. \textbf{Specific examples of the desired outcomes} can be listed for the given inputs~\cite{Shneiderman20}. \textbf{Negative requirements} with misuse, abuse and confuse cases can be described in the specification~\cite{Ebert19}.

\textbf{Requirements Verification and Validation: }
The ethical requirements should be specified \textbf{traceable} both backward to the stakeholders 
and forward into the design modules, code pieces
, and test cases. 
Requirements validation becomes an activity that needs
to be performed \textbf{continuously} during operation, which includes both monitoring and analysis of production data from the ethical perspective~\cite{Req_REW}. \textbf{Awareness of potential mismatches between training data and real-world data} is necessary to prevent the trained model from being unsuitable for its intended purpose. Model update or recalibration on new data is important for the trustworthiness of AI systems. The conditions of retraining need to be considered during the specification phase, such as time, frequency, data characteristics, user feedback. The \textbf{intended operating environment} should be understood as complete as possible. For example, the knowledge about the \textbf{input data} fed into the decision-making component of the AI system 
and the \textbf{data sources} used to obtain the input data 
should also be learnt~\cite{Steimers21}.

\subsubsection{Design}\noindent \par
Significant efforts have been put on ethical algorithms where ethical properties have been enabled via mathematical definitions, analysis and theoretical guarantees, such as fairness~\cite{mehrabi2019survey} and privacy~\cite{ji2014differential}. 
However, these algorithm focused mechanisms have limited theoretical heuristic and are confined to a small subset of ethical principles that can be easily translated to quantifiable ethical properties. Most of the time, these ethical-aware algorithms are too complicated to explain to less numeracy-equipped stakeholders~\cite{CDEI20} and hardly connected to the development processes.

There are different ways in the design process to reduce the ethical risk by managing: 1) \textbf{Frequency of occurrence}: The frequency of automatic decision-making by AI systems can be reduced. Instead, AI systems can make suggestions to human and ask for human's approval on the final decision; 2) \textbf{Consequence size}: The consequence size can be managed through deployment strategies, e.g. only deploying the new model version to a group of users; 3) \textbf{Consequence response}: This can be done through worst case analysis (e.g., FMEA), the resilient design to recover the state, or punishing through the incentive mechanism, or overriding the recommended decisions, or undoing the actions.

There has been emerging research on the design process that considers ethical principles. \textbf{Value sensitive design}\cite{VSD2015} is a design approach that incorporates human value into the whole design process. To drive the value sensitive design for responsible AI systems, \textbf{participatory co-design workshops} can be organized for developers and stakeholders~\cite{Bilstrup20} using different types of toolkits and methods, such as \textbf{card-based toolkits} \cite{Steven21,Shen21, Bilstrup20}, \textbf{ethical matrix}~\cite{muhlenbach2020}, \textbf{ethical hackathon}~\cite{Patel2019}, \textbf{user journeys}, \textbf{sketches}, \textbf{low-fidelity paper prototypes}, \textbf{high-fidelity clickable wireframes}~\cite{Stumpf2021}, \textbf{role play scenarios}, \textbf{table-top poll}, \textbf{low-fidelity storyboards}~\cite{Skinner20}, value sensitive algorithm design~\cite{Zhu18}. 

There are a variety of card-based toolkits that can be used in the design workshops. \textbf{Envisioning cards} are one of the most adopted card-based toolkits in practice, which help designers envision of the long term effect their systems will have on stakeholders.
~\cite{Steven21, Bilstrup20}. In addition to envisioning cards, there are other card-based toolkits that can be used in the value sensitive design workshops\cite{Shen21, Bilstrup20}, including \textbf{model cards} describing models with the inherent design trade-offs, \textbf{data cards} analyzing possible data sources, \textbf{people/persona cards} discussing potential stakeholders and their values and interests in the system, \textbf{criteria/checklist cards} considering different social and technical criteria, \textbf{ethics cards} reflecting on ethical implications of implementing an AI system, \textbf{situation cards} describing 
problematic situations, \textbf{inspiration cards} supporting participants to generate innovative and 
design concepts with new technologies. Some card-based toolkits are designed specifically for a type of applications, such as Tiles for designing IoT applications and PlutoAR for designing augmented reality (AR) applications~\cite{Bilstrup20}.

Modelling has been adapted to reflect ethical concerns and relevant design decisions, including using \textbf{SysML} to represent the architecture and describe the ethical aspects of AI systems~\cite{Takeda19}, designing \textbf{formal models} incorporating human values~\cite{Fish21}, using \textbf{ontologies} to represent the artifacts 
of AI systems to make them accountable~\cite{Naja21}, building up \textbf{ethical knowledge bases} recommending design paths considering ethical concerns~\cite{Sekiguchi20}, using \textbf{inductive logic programming} to codifying ethical principles~\cite{Anderson14}.

Before deploying AI systems in real-world, it is important to perform \textbf{system-level simulation} to understand the behavior of AI systems and evaluate ethical quality attributes in a cost-effective way. \textbf{ethical scenario simulation}~\cite{Singh21} is an effective simulation-driven design method. A \textbf{digital twin} simulates what is happening to an AI system running in the real world, which can help find unethical issues at run-time and improve the design.

\textbf{Trust factors} should be analyzed and considered during the design, such as capability of systems, availability of interface, personality of agent (e.g., voice embodiment, visual virtual figure or physical representation)~\cite{Wu21}. The four major \textbf{factors for trustworthiness-by-design} include data, algorithm, architecture, software~\cite{Liu20}.


Many efforts for explainable AI (XAI) user interface design have been spent on checklists~\cite{Liao20,liao21, larasati21}. The \textbf{question-driven checklists} are an effective way to understand the user needs, choices of XAI techniques, and XAI design factors~\cite{liao21}. The main factors for the XAI design include information included, information delivery, and interaction included~\cite{larasati21}. The questions can be classified into the following groups: input, output, performance (can be extended to ethical performance), how, why, why not, what if, how to be that, how to still be this~\cite{Liao20}. 
The conversational interface design can be experimented via a \textbf{wizard of oz} in which users interact with a system that the users believe to be autonomous but is actually being operated by an unseen human~\cite{jentzsch19}. 
There are ways to increase the level of human trust in AI systems through XAI user interface: i) \textbf{integrating human realism} (i.e. anthropomorphism), including appearance and other behavioural characteristics, into the interface design ~\cite{Luxton14}; ii) proactively informing users and the public data use information;  
iii) providing measurable benefits to users; iv) informing users capabilities, scope of use, and limitations of AI systems; v) informing users credentials 
of AI systems and operators; vi) explanations of data, algorithm, models, system decisions and behaviours; vii) considering explanation audiences' explainability needs across the AI project lifecycle.

\subsubsection{Implementation}
\textbf{Ethical implementation standards} are collections of implementation rules from the ethical perspective, including communication methods (i.e., documentation), programming languages, coding, interface, and tooling~\cite{SWEBOK}.
Developers need to follow the communication standards to continuously maintain high-quality, up-to-date code \textbf{documentation} that covers both AI components and non-AI component. The \textbf{author of each line and who made changes} need to be recorded and maintained in a repository to enable accountability and traceability. 
\textbf{Code review} can be conducted following the predefined ethical standards. 
\textbf{Ethical principles and metrics} need to be defined and added into the developments tools to automate ethical quality checks. Initial attempts have been made on responsible AI tooling support, mainly on algorithm-level rather than system-level, such as Google
\footnote{\url{https://cloud.google.com/responsible-ai}} and Microsoft
\footnote{\url{https://www.microsoft.com/en-us/ai/responsible-ai-resources}}. In particular, there are many industry fairness toolkits, such as IBM's AI Fairness 360, Google's Fairness Indicators, Microsoft's Fairlearn, and UChicago's Aequitas.

There may be ethical quality issues with APIs (e.g., data privacy breaches or fairness issues). Thus, \textbf{ethical compliance checking for APIs} is needed to detect if any ethics violation exists. Ethical knowledge graphs can be built based on the ethical principles and guidelines (e.g. privacy knowledge graph based on GDPR~\cite{gdpr}) to automatically examine whether APIs are compliant with regulations for AI ethics. Call graph might also be needed for code analysis as there might be interactions between different APIs.

\textbf{Construction with reuse} means to develop responsible AI systems with the use of existing AI assets, e.g. from an organizational repository or an open-source platform. A \textbf{marketplace} can be built up to trade the reusable AI assets, including component code, models, and datasets. Blockchain can be adopted to design an immutable and transparent marketplace enabling the auction-based trading for AI assets and material assets (e.g., cloud resources)~\cite{Nicolas22}. To ensure the ethical quality, \textbf{credentials} can be bound with the AI assets or developers, which can also be supported by blockchain platforms. If different frameworks are used in model migration, guidelines and tooling support (such as pytorch2keras\footnote{\url{https://github.com/gmalivenko/pytorch2keras}}) are needed to \textbf{automatically migrate a model} from one framework to another. \textbf{Glue code} may be used to integrate AI components with non-AI components to eliminate incompatibility since there are various inputs and outputs for the AI component~\footnote{\url{https://insights.sei.cmu.edu/blog/software-engineering-for-machine-learning-characterizing-and-detecting-mismatch-in-machine-learning-systems/}}.

\subsubsection{Verification and Validation}\noindent \par
Verification and validation are used together for checking whether an AI system meets the requirements described in the specification and fulfills its intended purpose in a responsible way. \textbf{Ethical acceptance testing} (e.g. \textbf{Bias testing}) can be designed for detecting the ethics-related design flaws ~\cite{Chattopadhyay21} and verifying the ethical requirements (e.g. whether the data pipeline has appropriate privacy control, fairness testing for training/validation data). The \textbf{behaviour of the AI system should be quantified} by the acceptance testing and the \textbf{acceptance criteria} for each of the ethical principles should be defined in a testable way. A \textbf{testing leader} may be appointed to lead the testing for each ethics principle. For example, When bias detected at runtime, the monitoring reports are returned to the bias testing leader~\cite{Shneiderman20}. An \textbf{ethical scoring system} using a set of actionable tests can be used to measure how ready for production a given AI system is from the ethics perspective. When certifying ethical aspects of an AI system/component, \textbf{benchmark testing} may be needed. \textbf{Formal verification} can be used to prove that a system matches its ethical requirements through a comprehensive mathematical analysis~\cite{Yap21, Dennis21}.

A collection of test cases with expected results should be maintained to detect possible ethical failures in a wide variety of extreme situations. New test cases need to be added when there is a new requirement added or the operation context changes~\cite{Shneiderman20}. All the test cases for verification and validation should \textbf{pass the ethics assessment}. The \textbf{history of testing should be recorded and tracked}, such as how and by whom the ethical issues were fixed.

The traditional testing techniques can be adapted for testing AI systems. \textbf{Unit testing} can be performed for both AI components and non-AI components according to the specification (including model-level specification). Interactions between AI system components, particularly in between AI components (i.e. for AI pipeline) and between AI components and non-AI components, need to be verified through \textbf{incremental integration testing} along the development process~\cite{Ebert19}. \textbf{Infrastructure testing} is also part of the integration testing. \textbf{Sanity testing} is performed after the software build to ensure that the code changes introduced are working aligned with ethical principles. \textbf{Usability testing}~\cite{Shneiderman20} measures stakeholder performance and satisfaction and is essential to ensure the systems is easy to use and does what users and indirect stakeholders expect in terms of responsible AI. \textbf{Data tests} need to be performed to check whether the serving data is the data we expect in the operating environment, e.g., check the inputs for each variable match what the model expects. \textbf{Skew tests} checks how representative the training data is of the serving data, e.g., through the percentage of missing data in the serving data compared to the training data. AI4SE4AI can be applied to automate the testing through \textbf{AI testing} and/or \textbf{cognitive testing}~\cite{Ebert19}. 

Both \textbf{failure mode and effects analysis (FMEA)} and \textbf{fault tree analysis (FTA)} are tools to understand ethical failures and risk of AI systems~\cite{Ebert19}. \textbf{Assurance cases}~\cite{Yanagisawa21} (e.g. safety case~\cite{Gauerhof20}) provide evidence and arguments that support claims about ethical properties or behaviors of AI systems. There are four levels of evaluation assurance~\cite{Yap21}, including basic disclosure, tested claims, active testing, and formal verification.

\textbf{Scoring tools} allow the team to measure the level of trustworthiness~\cite{Jordan19} or trust (e.g., trustworthiness/trust score) by assessing the outcomes of AI systems based on the contextual data or capturing users' experiences with AI features. \textbf{Artificial intelligent quotient}~\cite{tschopp2018trust} tracks a conversational AI system's level of competence and capabilities over time (e.g. knowledge, language, reasoning, creative and critical thinking, working memory).
The agents' \textbf{humanness factors} (e.g. speaking and listening) can help improve the trust in conversational AI systems~\cite{hu21}.

\subsubsection{Operations}\noindent \par
Given the continual learning of AI systems based on new data and the higher degree of uncertainty and risks associated with the autonomy of the AI systems, there is a strong desire for deployment strategies and continuous validation of responsible AI requirements. The deployment strategies include \textbf{phased deployment} (i.e., deploying AI systems for a subset group of users initially to reduce ethical risk), \textbf{homogeneous redundancy}
, etc.

The existing work on the \textbf{continuous monitoring and validation} mainly focuses on  the AI system outputs (e.g., performance metrics - accuracy, precision, and recall ) rather than the \textbf{outcomes} (i.e. whether the AI system behaves and make decisions responsibly)~\cite{cmu_human_centered_ai}. 
With different \textbf{context} data (e.g., users, traffic, weather) in operation, AI systems are continuously evolving to address ethical risks. 
The current practice for ethical risk assessment is often one-off type of risk assessment (e.g. Five Safes Framework\footnote{\url{https://www.aihw.gov.au/about-our-data/data-governance/the-five-safes-framework}}). \textbf{Ethical risk assessment is expected to be dynamic, adaptive, and extensible} for different context e.g., culture. 
.
\textbf{Version-based feedback} (e.g., 
ethical violation) should be reported to the development team and other stakeholders continuously. \textbf{Incentives} can be applied to encourage the ethical outcomes of AI systems in terms of both decisions and behaviors. The time and frequency of validation and conditions of necessary retraining should be predefined~\cite{Req_REW}.

An AI system usually involves co-evolution of data, model, code, and configurations. \textbf{Data/model/code/configuration co-versioning} traces exactly what datasets and configuration parameters were used to train and evaluate the model. 
Co-evolution may also happen to AI components and non-AI components, thus co-versioning is required to facilitate the maintenance and communication between the AI component development team and non-AI component team.

\textbf{Bill of materials} (BOM) ~\cite{SBOM,barclay19} enables the transparency and accountability by keeping a formal record of the supply chain details of the components used in building an AI system, such as component name, version, supplier, dependency relationship, author and timestamp. 
\textbf{Ethical audit trail} records every step of AI systems from the ethics perspective. Developers need to consider trade-offs between accountability and performance when making design decisions on what date is placed on blockchain/cloud. \textbf{Accountability knowledge graphs} support capturing accountability information throughout the lifecycle of AI systems and auditing them programmatically. To build such knowledge graphs, \textbf{ontology} can be used to describe and model accountability information~\cite{Naja21, markovic2021accountability}.

\subsection{Research Challenges}
\textbf{Integrating the software development process with the AI model development pipeline for ethical principles.} 
There is a methodological gap between the AI model development pipeline that produces the AI model and software development process that develops the non-AI components and AI components embedding AI model~\cite{martinez2021developing}. Given the AI model pipeline is more experimental with still limited methodological support, integrating the AI model pipeline into the agile software development process needs better understanding of artifacts, activities and roles involved. Although initial tooling attempts have been made on the integration (such as Microsoft Team Data Science Process\footnote{\url{https://docs.microsoft.com/en-us/azure/architecture/data-science-process/overview}} and IBM Watson Studio\footnote{\url{https://www.ibm.com/cloud/watson-studio}}), the integrated AI system development process still need to be standardized and supported by software tools taking into account responsible AI principles.

\textbf{Capturing ethical principles by requirement engineering.} 
Compared with traditional software and general responsible software, AI systems also need to consider requirements about models, training data, system autonomy oversight and may emphasize certain ethical requirements more due to AI-based autonomous and potentially opaque behaviour and decision making. In particular, some of the ethical principles are hard to define and quantify. To make the ethical principles be captured by requirements engineering for AI systems, the community should put further efforts on requirements engineering methods and provide a concrete guidance on requirement engineering for responsible AI systems.


\textbf{Designing AI systems for improving both trustworthiness and trust.}
Trustworthiness is the ability of an AI system to meet ethical principles, while trust is users' subjective estimates of the trustworthiness
\cite{zhu2021ai}. Trust in AI systems involves trust duality that includes trust in providers and trust in technologies that can be further classified into trust in an AI technology and a base technology (e.g. vehicle) ~\cite{Renner21}. Both trust in a provider and trust in a technology need to be considered in parallel. A user's trust in an AI system may have a significant gap compared to the AI system's inherent trustworthiness, i.e., a user underestimates or overestimates a system's trustworthiness and has inadequate or excessive trust into the system. More efforts need to be made on the design to improve both trustworthiness and trust in AI systems. Process and product (i.e. system perspective) mechanisms can be leveraged to achieve trustworthiness for different ethical principles, whereas process and product evidence need to be offered for different types of trusters in a proper communication way to drive trust. These will help close the gap between their subjective estimation and the system's more objective trustworthiness. Instead of focusing on verifiable product trustworthiness via mathematical algorithm-level guarantees, researchers need to systematically explore a broader variety of mechanisms in system-level product design and development processes to improve both trustworthiness and trust. 

\textbf{Continuously monitoring and validating the outcomes of AI systems against responsible AI requirements.}
There are two forms of AI system development defined in ~\cite{jan2019}: requirement-driven development and outcome-driven development. The latter is about the real-world operation and outcome of AI systems. Seamless integration of requirement-driven development and outcome-driven development requires understanding the unique characteristics of AI systems. The development of AI systems is a continual and iteration process. Validation of outcomes (i.e. whether the system provides the intended benefits and behave appropriately given the situation) is required for AI systems rather than just outputs (e.g. precision, accuracy and recall).
Also, since all principles need to be instantiated, it is necessary to make risk assessment and management extensible , adaptive, and dynamic for different context, with guided extension points. For example, some principle might be automatically instantiable for different culture context and extended following guided extension points. 
Further work is needed on MLOps for continuously monitoring and validating the outcomes of AI systems against the responsible AI requirements.



\begin{figure*}[t]
	\centering
	\includegraphics[width=\textwidth]{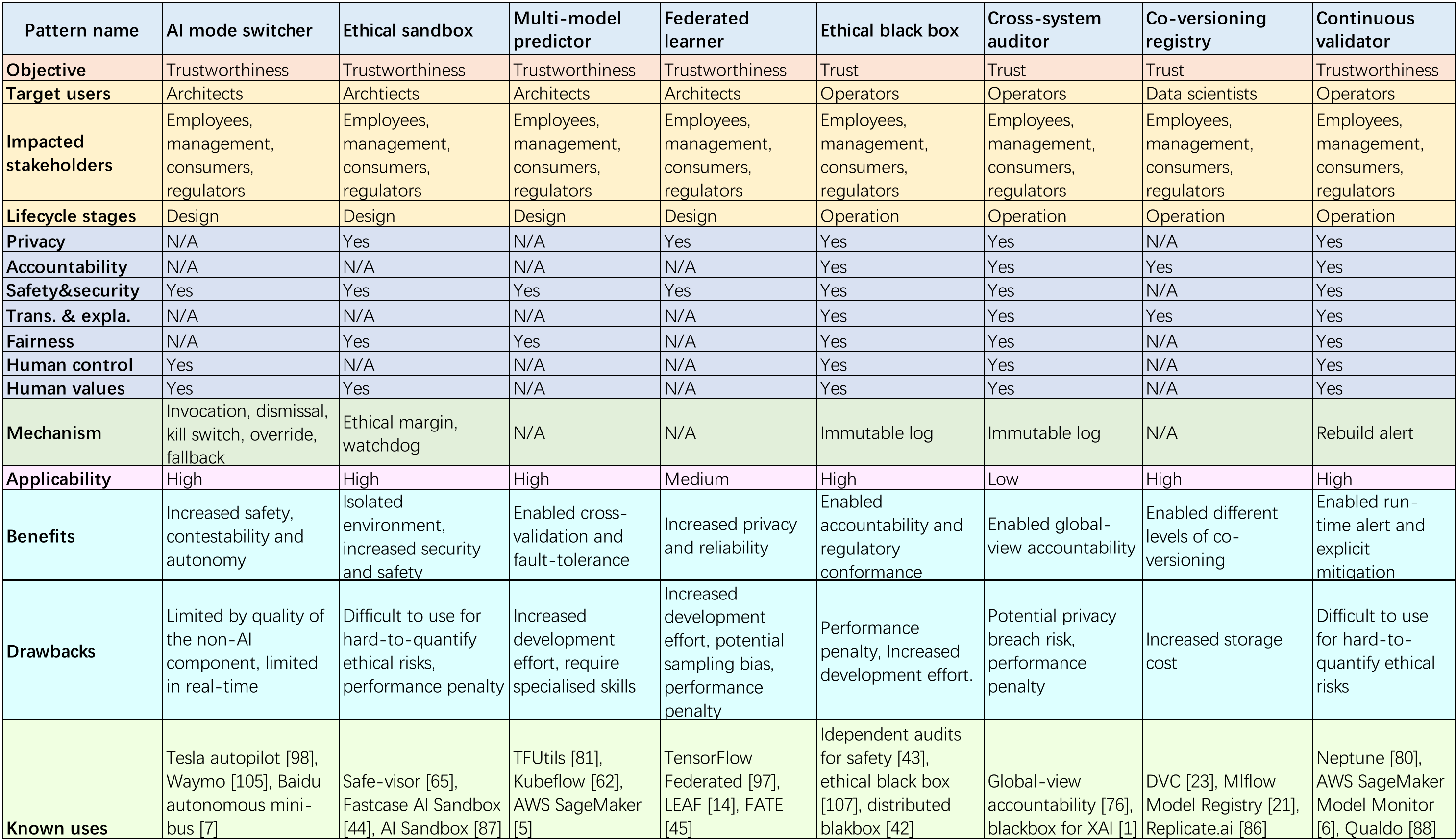}
	\caption{Architectural patterns for responsible-AI-by-design.}
	\label{fig:patterns}
\end{figure*}

\section{System Perspective}

\subsection{Current State}

\subsubsection{Architectural style and patterns}\noindent \par
Responsible-AI-by-design architectural style can be defined by a set of design principles and patterns. Some design principles include \textbf{co-architecting of AI components and non-AI components, minimum complexity, design with reuse}.
The architecture of AI systems includes AI components that produce and embed the AI models and non-AI components that uses the outputs of AI components for overall system functionalities. As the AI components are often iteratively experimented by data scientists/engineers who are not familiar with software engineering, \textbf{co-architecting of the AI components and the non-AI components} can ensure the seamless integration of the two types of components and consideration of both system-level and model-level (ethical) requirements when making design decisions. Scenario-based evaluation methods can be adapted to evaluate the architecture of AI systems. 
Developers should follow the principle of \textbf{minimum complexity} that includes both the design of AI systems (e.g. software/system architecture, whether adopting AI or not, selection of AI techniques) and the future operating environment~\cite{Steimers21}. 
One principle is \textbf{design with reuse}. AI system components, particularly the AI components for producing AI models, should be evaluated against responsible AI requirements and \textbf{reused} as much as possible to improve productivity~\cite{Steimers21}. Ethical certificates may be needed for reusing the AI components developed by the third party.

Fig.~\ref{fig:patterns} lists a set of architectural patterns for responsible-AI-by-design, which could be embedded as the product features of AI products. Adopting AI or not can be considered as a major architectural design decision when designing a software system. For example, \textbf{AI mode switcher} offers users efficient \textbf{invocation and dismissal} mechanisms for activating or deactivating the AI component when needed~\cite{vassilakopoulou2020sociotechnical, Tesla, Waymo, Apollo}. \textbf{Kill switch} is a special type of invocation mechanism which immediately shuts down the AI component and its negative effects. The decisions made by the AI component can be executed automatically or reviewed by a human expert in critical situations. The human expert serves to approve or \textbf{override} the decisions. Human intervention can also happen after acting the AI decision through the \textbf{fallback} mechanism that reverses the system back to the state before executing the AI decision. A \textbf{built-in guard} ensures that the AI component is only activated within the predefined conditions (such as domain of use, boundaries of competence). Users can ask questions or report complaints/failures/near misses through a \textbf{recourse channel}. 

\textbf{Ethical sandbox} can be applied to run the AI component separately (e.g. from untrusted third parties)~\cite{Fastcase, AI_Sandbox}. Maximal tolerable probability of violating the responsible AI requirements should be defined as \textbf{ethical margin} for the sandbox~\cite{lavaei2021towards}. A \textbf{watch dog} can be used to limit the execution time of the AI component to reduce the ethical risk~\cite{serban2019designing}.
\textbf{Multi-model predictor} employs two or more models to perform the same task~\cite{TFUtils_Multi-Model_Training, Kubeflow, Amazon_SageMaker}. This pattern can improve the reliability by deploying different models under different context (e.g., different regions) and enabling fault-tolerance by cross-validating ethical requirements for a single decision (e.g., only accepts the same results from the employed models)~\cite{dai2021more, nafreen2020architecture}.
\textbf{Federated learner} preserves the data privacy by training models locally on the client devices and formulating a global model on a central server based on the local model updates~\cite{TensorFlow_Federated, LEAF, FATE}.
\textbf{Decentralized learning} is an alternative to federated learning, which uses blockchain to remove the single point of failure and coordinate the learning process in a fully decentralized way.

In the event of negative outcomes, the responsible humans can be 
identified by an \textbf{ethical black box} for accountability~\cite{falco2021governing}. The ethical black box continuously records  sensor data, internal status data, decisions, behaviors (both system and operator) and effects~\cite{winfield2017case}. All of these data need to be kept as evidence with the timestamp and location data using an \textbf{immutable log} (e.g. using blockchain)\cite{falco2020distributedblack}. 
\textbf{Cross-system auditor} provides global-view accountability by finding discrepancies among the data collected from a set of AI systems and identifying liability when negative events occur~\cite{miguel2021putting, Black_Box}. 
\textbf{Co-versioning registry} can be applied to ensure accountability in two cases: 1) co-versioning of AI components and non-AI components; 2) co-versioning the artifacts within the AI components, i.e., data, model, code, and configuration~\cite{DVC, Mlflow_Model_Registry, Replicate}.
\textbf{Continuous validator} supports continuous ethical risk assessment by monitoring and validating the predefined ethical metrics~\cite{Neptune, SageMaker, Qualdo}. The time and frequency of validation should be configured within the continuous validator. \textbf{Version-based feedback} and \textbf{rebuild alert} should be sent when the predefined conditions are met. \textbf{Incentive mechanisms} can be designed to reward/punish the ethical/unethical behavior or decisions of AI systems.

\subsubsection{System-Level Techniques}
There are many system-level techniques, which could be embedded as components:
\begin{itemize}
  \item Fairness: demographic parity, data augmentation, weighted data sampling, re-sampling, re-weighting, swapping labels, removing dependencies, equalized odds checking;
  \item Privacy: data sanitizing, federated learning, decentralized learning, differential privacy, secure multiparty computation, homographic learning, fog computing;
  \item Safety and security: sandboxing, trusted execution environments; safety margin; data representativeness checking, approximate computing;
  \item Explainability: global explanations, local explanations, feature-based explanations (contrastive, counterfactual, rule-based explanations), exemplar-based explanations, post hoc explanations for black box models,	prospective explanations, surrogate models (LIME, SHAP, PyExplainer~\cite{Chanathip21}), what-if, confidence scores.
\end{itemize}

\subsection{Research Challenges}

\textbf{Defining architectural style for responsible-AI-by-design.}
An AI system consists of AI components and non-AI components
that are interconnected and work together to achieve the system's objective~\cite{lu2021software}.
An AI model needs to be trained and integrated into the inference component of the system to perform the required functions. Both the AI pipeline components and the inference component can be considered as AI components.
Combining AI and non-AI components may create new emergent behavior and dynamics.
Therefore, ethics need to be considered at system-level,
including AI components, non-AI components and their connections.
For example, the new data gathered by the data collector need to be fed into the model trainer, whereas the effect of actions decided by the AI model may change the behavior of non-AI components and could be collected through the feedback component built into the AI system. One way to achieve responsible-AI-by-design is to define an architectural style through a set of architectural design principles taking into account ethical principles. Additional principles need to be explored in addition to the initial attempt.

\textbf{Dealing with ethical requirement trade-offs using design patterns.}
There are trade-offs between functional requirements and ethical principles or in between some of the ethics principles. Privacy and utility are often conflicting. For example, to fulfill the privacy requirements, the datasets can be de-identified and aggregated so that individuals cannot be uniquely identified~\cite{ahuja2020}, which may lead to worse distributional properties and affect the reliability. Requirement inconsistency also happens to accountability and privacy. For example, when particular activities are not compliant to some regulatory standards, we need to find out 
where this happened and who to blame. In this case, there might be an issue about data privacy. 
The current practice dealing with conflicting ethics principles is usually the developers following one principle while overriding the other rather than building balanced trade-offs with stakeholders making the ultimate value and risk call. Patterns can be used to deal with the system-level trade-offs among conflicting responsible AI principles and other requirements in an inclusive manner. A pattern catalogue is needed to provide concrete guidance on how to design responsible AI systems.


\section{Conclusion}
To operationalize responsible AI, we present a research roadmap on software engineering for responsible AI. The findings could be organised as operationalized guidelines for the stakeholders of AI systems (e.g. regulators, management, developers). Some findings could be framed as tools (e.g. ethical risk assessment) or embedded as product features of AI systems to reduce the ethical risk and unlock the market where there is currently little trust (e.g. ethical black box). Although the major industry solutions (such as model cards) have been identified in our study through Google scholar and the snowballing process, we plan to do an industry landscape that will cover the complete state of the practice. 

\input{main.bbl}



\end{document}